%% file: main.tex
\documentclass[runningheads]{llncs}
\usepackage{draftwatermark}
\SetWatermarkLightness{0.8}
\SetWatermarkText{PREPRINT}
\SetWatermarkScale{1.5}

\include{macros}

\begin{document}
\lstset{style=prolog-style}

\title{Towards Continuous Assurance Case Creation for ADS with the Evidential Tool Bus}
\author
{
Lev Sorokin\thanks{The authors contributed equally to this paper.}
 \inst{1} \and
Radouane Bouchekir$^*$ \inst{1} \and
Tewodros A. Beyene$^*$ \inst{1} \and
Brian Hsuan-Cheng Liao \inst{2}
\and
Adam Molin \inst{2}
}

\institute{fortiss GmbH, An-Institut Technische Universität München, Guerickestraße 25, 80805 München, Germany \\ \email{ \{sorokin, bouchekir, beyene\}@fortiss.org} 
\and
DENSO AUTOMOTIVE Deutschland GmbH, Freisinger Str. 21, 85386 Eching, Germany  \\
\email{ \{h.liao, a.molin\}@eu.denso.com}
}

\authorrunning{Lev Sorokin, Radouane Bouchekir, Tewodros A. Beyene et al.}
\titlerunning{Assurance Case Maintenance using a Tool Integration Framework}
\maketitle

\input{Sections/S0_Abstract}

\input{Sections/S1_Intro}

\input{Sections/S4_UCAVP}
\input{Sections/SX_Application}

\input{Sections/SX_Lessons}
\input{Sections/S2_RelatedWork}

\input{Sections/S6_Results_Conclusion}

\section*{Acknowledgments}
\includegraphics[scale=0.02]{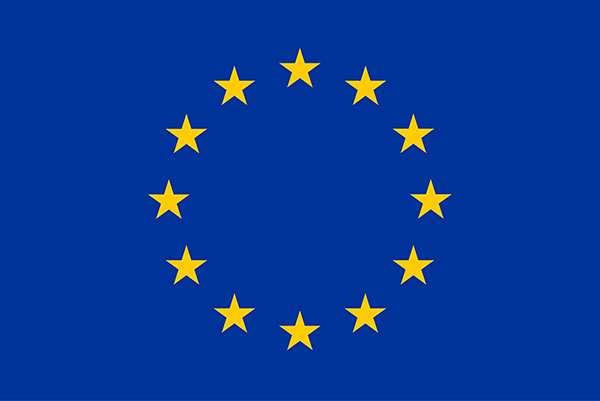} 
This work is part of FOCETA project that has received funding from the European Union’s Horizon 2020 research and innovation program under grant agreement N. 956123. 


\bibliographystyle{splncs04} 
\bibliography{paper.bib}

\end{document}

%% file: macros.tex
\usepackage[square,sort&compress,comma,numbers]{natbib}

\usepackage{todonotes}
\usepackage{listings}
\usepackage{amsmath,amssymb,amsfonts}
\usepackage{algorithmic}
\usepackage{graphicx}
\usepackage{textcomp}
\usepackage{hyperref}
\usepackage{xspace}

\usepackage{lipsum}
\usepackage{multicol}

\include{lst_style}

\lstdefinestyle{javaStyle}{
  language=Java,
  basicstyle=\small\scriptsize,
  keywordstyle=\color{blue}\bfseries,
  commentstyle=\color{green!60!black},
  stringstyle=\color{orange},
  numbers=left,
  numberstyle=\tiny,
  stepnumber=1,
  numbersep=5pt,
  backgroundcolor=\color{gray!10},
  frame=single,
  rulecolor=\color{black!30},
  breaklines=true,
  breakatwhitespace=true,
  tabsize=4
}
\lstdefinestyle{smallFrame}{
  backgroundcolor=\color{backcolour}, commentstyle=\color{codegreen},
  keywordstyle=\color{magenta},
  numberstyle=\tiny\color{codegray},
  stringstyle=\color{codepurple},
  basicstyle=\ttfamily\scriptsize,
  breakatwhitespace=false,         
  breaklines=true,                 
  captionpos=b,                    
  keepspaces=true,                 
  numbers=left,                    
  numbersep=5pt,                  
  showspaces=false,                
  showstringspaces=false,
  showtabs=false,                  
  tabsize=2
}

\newcommand{\AVP}{\texttt{AVP}\xspace}
\newcommand{\ETB}{\texttt{ETB}\xspace}
\newcommand{\SUA}{\texttt{SUA}\xspace}

\newcommand{\goal}{\texttt{G}}

\usepackage{color,soul}

\usepackage{tikz}
\usetikzlibrary{patterns}
\usepackage{hyperref}
\long\def\omitthis#1{\relax}

\usepackage{textcomp}
\usepackage[export]{adjustbox}
\usepackage{listings}
\usepackage{wrapfig}
\usepackage{sansmath}
\usepackage{microtype}

\renewcommand{\footnotesize}{\scriptsize}

\usepackage{subcaption}
\usepackage{caption}



%% file: lst_style.tex
\usepackage{xcolor}

\definecolor{codegreen}{rgb}{0,0.6,0}
\definecolor{codegray}{rgb}{0.5,0.5,0.5}
\definecolor{codepurple}{rgb}{0.58,0,0.82}
\definecolor{backcolour}{rgb}{0.95,0.95,0.92}

\lstdefinestyle{mystyle}{
    backgroundcolor=\color{backcolour},   
    commentstyle=\color{codegreen},
    keywordstyle=\color{magenta},
    numberstyle=\tiny\color{codegray},
    stringstyle=\color{codepurple},
    basicstyle=\ttfamily\scriptsize,
    breakatwhitespace=false,         
    breaklines=true,                 
    captionpos=b,                    
    keepspaces=true,                 
    numbers=left,                    
    numbersep=5pt,                  
    showspaces=false,                
    showstringspaces=false,
    showtabs=false,                  
    tabsize=2
}

\makeatletter

\newif\ifpredicate@prolog@
\newif\ifwithinparens@prolog@

\lst@SaveOutputDef{`_}\underscore@prolog

\newcount\currentchar@prolog

\newcommand\@testChar@prolog%
{%
  \ifnum\lst@mode=\lst@Pmode%
    \detectTypeAndHighlight@prolog%
  \else
    \ifwithinparens@prolog@%
      \detectTypeAndHighlight@prolog%
    \fi
  \fi
  \global\predicate@prolog@false%
}

\newcommand\detectTypeAndHighlight@prolog
{%
  \def\lst@thestyle{\PrologAtomStyle}%
  \ifpredicate@prolog@%
    \def\lst@thestyle{\PrologPredicateStyle}%
  \else
    \expandafter\splitfirstchar@prolog\expandafter{\the\lst@token}%
    \expandafter\ifx\@testChar@prolog\underscore@prolog%
      \ifnum\lst@length=1%
        \let\lst@thestyle\PrologAnonymVarStyle%
      \else
        \let\lst@thestyle\PrologVarStyle%
      \fi
    \else
      \currentchar@prolog=65
      \loop
        \expandafter\ifnum\expandafter`\@testChar@prolog=\currentchar@prolog%
          \let\lst@thestyle\PrologVarStyle%
          \let\iterate\relax
        \fi
        \advance \currentchar@prolog by 1
        \unless\ifnum\currentchar@prolog>90
      \repeat
    \fi
  \fi
}
\newcommand\splitfirstchar@prolog{}
\def\splitfirstchar@prolog#1{\@splitfirstchar@prolog#1\relax}
\newcommand\@splitfirstchar@prolog{}
\def\@splitfirstchar@prolog#1#2\relax{\def\@testChar@prolog{#1}}

\def\beginlstdelim#1#2%
{%
  \def\endlstdelim{\PrologOtherStyle #2\egroup}%
  {\PrologOtherStyle #1}%
  \global\predicate@prolog@false%
  \withinparens@prolog@true%
}

\newcommand\lang@prolog{Prolog-pretty}
\expandafter\lst@NormedDef\expandafter\normlang@prolog%
  \expandafter{\lang@prolog}

\expandafter\expandafter\expandafter\lstdefinelanguage\expandafter%
{\lang@prolog}
{%
  language            = Prolog,
  keywords            = {},      
  showstringspaces    = false,
  alsoletter          = (,
  moredelim           = **[is][\beginlstdelim{(}{)}]{(}{)},
  MoreSelectCharTable =
    \lst@DefSaveDef{`(}\opparen@prolog{\global\predicate@prolog@true\opparen@prolog},
}

\newcommand\@ddedToOutput@prolog\relax
\lst@AddToHook{Output}{\@ddedToOutput@prolog}

\lst@AddToHook{PreInit}
{%
  \ifx\lst@language\normlang@prolog%
    \let\@ddedToOutput@prolog\@testChar@prolog%
  \fi
}

\lst@AddToHook{DeInit}{\renewcommand\@ddedToOutput@prolog{}}

\makeatother
\definecolor{blue3F6696}{HTML}{3F6696}
\definecolor{blueADC1D7}{HTML}{ADC1D7}
\definecolor{blueD3E0EE}{HTML}{D3E0EE}
\definecolor{darkyellow}{rgb}{0.75,0.75,0.0}
\definecolor{lightgray}{rgb}{0.8,0.8,0.8}
\definecolor{orange}{rgb}{1.0,0.5,0.0}
\definecolor{verylightgray}{rgb}{0.95,0.95,0.95}
\definecolor{green-verystrong}{HTML}{4F6228}
\definecolor{green-strong}{HTML}{76923C}
\definecolor{green-medium}{HTML}{C2D99B}
\definecolor{green-low}{HTML}{D6E3BC}
\definecolor{green-no}{HTML}{FFFFFF}

\newcommand\PrologPredicateStyle{}
\newcommand\PrologVarStyle{}
\newcommand\PrologAnonymVarStyle{}
\newcommand\PrologAtomStyle{}
\newcommand\PrologOtherStyle{}
\newcommand\PrologCommentStyle{}

\definecolor{PrologPredicate}{RGB}{000,031,255}
\definecolor{PrologVar}      {RGB}{024,021,125}
\definecolor{PrologAnonymVar}{RGB}{000,127,000}
\definecolor{PrologAtom}     {RGB}{186,032,032}
\definecolor{PrologComment}  {RGB}{063,128,127}
\definecolor{PrologOther}    {RGB}{000,000,000}

\renewcommand\PrologPredicateStyle{\color{PrologPredicate}}
\renewcommand\PrologVarStyle{\color{PrologVar}}
\renewcommand\PrologAnonymVarStyle{\color{PrologAnonymVar}}
\renewcommand\PrologAtomStyle{\color{PrologAtom}}
\renewcommand\PrologCommentStyle{\itshape\color{PrologComment}}
\renewcommand\PrologOtherStyle{\color{PrologOther}}

\lstdefinestyle{prolog-style}
{
  language     = Prolog-pretty,
  upquote      = true,
  basicstyle=\ttfamily\fontsize{9pt}{10pt}\selectfont,
  breakatwhitespace=false,   
  stringstyle  = \PrologAtomStyle,
  commentstyle = \PrologCommentStyle,
  literate     =
    {:-}{{\PrologOtherStyle :-}}2
    {,}{{\PrologOtherStyle ,}}1
    {.}{{\PrologOtherStyle .}}1
}

\lstdefinestyle{prolog-style-small}
{
  language     = Prolog-pretty,
  upquote      = true,
  basicstyle=\ttfamily\scriptsize,
  breakatwhitespace=false,   
  stringstyle  = \PrologAtomStyle,
  commentstyle = \PrologCommentStyle,
  literate     =
    {:-}{{\PrologOtherStyle :-}}2
    {,}{{\PrologOtherStyle ,}}1
    {.}{{\PrologOtherStyle .}}1
}

%% file: Sections/S0_Abstract.tex
\begin{abstract}

An assurance case has become an integral component for the certification of safety-critical systems. 
While manually defining assurance case patterns can be not avoided, system-specific instantiations of assurance case patterns are both costly and time-consuming. It becomes especially complex to maintain an assurance case for a system when the requirements of the System-Under-Assurance change, or an assurance claim becomes invalid due to, e.g., degradation of a systems' component, as common when deploying learning-enabled components. 

In this paper, we report on our preliminary experience leveraging the tool integration framework Evidential Tool Bus (\ETB) for the construction and continuous maintenance of an assurance case from a predefined assurance case pattern. 
Specifically, we demonstrate the assurance process on an industrial Automated Valet Parking system from the automotive domain. We present the formalization of the provided assurance case pattern in the \ETB processable logical specification language of workflows. 
Our findings, show that \ETB is able to create and maintain evidence required for the construction of an assurance case.

\end{abstract}

%% file: Sections/S1_Intro.tex
\section{Introduction}




Assurance cases are an integral component of the certification process of safety-critical systems. They are based on a goal-oriented paradigm, which places greater emphasis on explicitly stating safety claims, and supplying an argument along with assurance evidence that needs to be generated to support these claims \cite{bishop2000methodology} \cite{hawkins2021guidance}. In general, such structured arguments and evidence are captured in the form of safety cases, i.e., a comprehensive, defensible, and valid justification of the safety of a system for a given application in a defined operating environment. 

The current practice of developing assurance cases is that safety engineers specify manually the arguments that connect higher-level safety claims to low-level assurance evidence. This practice of developing assurance cases is both highly expensive and labor-intensive, as it involves the extensive generation and meticulous maintenance of a substantial amount of evidence. In particular, assurance case development faces multiple challenges encompassing automation, tool integration, assurance distribution, and assurance maintenance. Although the manual definition of assurance case patterns may be unavoidable, the instantiation of system-specific assurance case patterns can be automated through the creation of automated assurance workflows. This automation not only reduces costs but also enhances efficiency by the automatic generation of assurance evidence. Tool integration presents another challenge, as diverse tools used in different phases of assurance case development must be integrated to generate assurance evidence. Yet, certification standards across various domains, such as DO-178C \footnote{Software Considerations in Airborne Systems and Equipment Certification} in avionics and ISO 26262 \footnote{ISO 26262 Road vehicles Functional safety.} in automotive, encourage the use of complementary tools when a single tool is not sufficient for a given test, analysis, or verification activity. The distribution of assurance across various stakeholders, system components, and platforms adds complexity, requiring careful management to ensure consistency and alignment with overarching assurance goals. In addition,  continuous maintenance is required, particularly for systems powered by learning-enabled components (LECs), as updates to assurance artifacts, e.g., data and models, occur frequently. This necessitates an efficient maintenance procedure that monitors all claims and assurance evidence affected by these changes. Moreover, such a procedure should aim to minimize maintenance costs by selectively and incrementally updating only the portions of the assurance case impacted by these changes. 

In this paper, we report on our experience in the development of an assurance case for an industrial Automated Valet Parking (\AVP) system from the automotive domain. Our focus is on utilizing \ETB \footnote{\url{https://git.fortiss.org/etb2/etb2}}\cite{cruanes2013tool} to address the previously mentioned challenges. Specifically, the automation of evidence generation and claims maintenance. In that regard, we illustrate the steps involved in creating an assurance case for the AVP system. This encompasses the assurance case pattern creation, formalization of the predefined assurance patterns, and considerations related to assurance distribution and maintenance. Additionally, we highlight the lessons learned through our experience leveraging an automated tool-chain for assurance case generation.

The rest of the paper is structured as follows. 
Section~\ref{sec:AVPUC} presents the \AVP case study. 
In section~\ref{sec:application}, we describe the assurance case development for the \AVP system and the use of the framework \ETB for assurance case generation. In Section~\ref{sec:discussion} we discuss the lessons learned from our study and present in Section~\ref{sec:related} related work. Finally, we provide concluding remarks and pointers to future work in Section~\ref{Sec:Conclusion}.

%% file: Sections/S4_UCAVP.tex
\section{The AVP Case Study}
\label{sec:AVPUC}


This section presents our \AVP case study, which is based on an \AVP System developed in the FOCETA project\footnote{\url{https://www.foceta-project.eu/}}. We use the \AVP system to demonstrate the usage of \ETB for the construction and maintenance of an assurance case.
%
%




\textbf{System Description.} The \AVP System is a feature added to a car that allows to autonomously park the car in an empty parking spot \cite{BoschAVP}. In particular, the car is dropped off by the driver in a designated zone, where \AVP takes over the car, computes the trajectory to a free parking spot and parks the vehicle in the spot.

\begin{wrapfigure}{R}{0.56 \textwidth}
 \centering
    \includegraphics[scale=0.27]{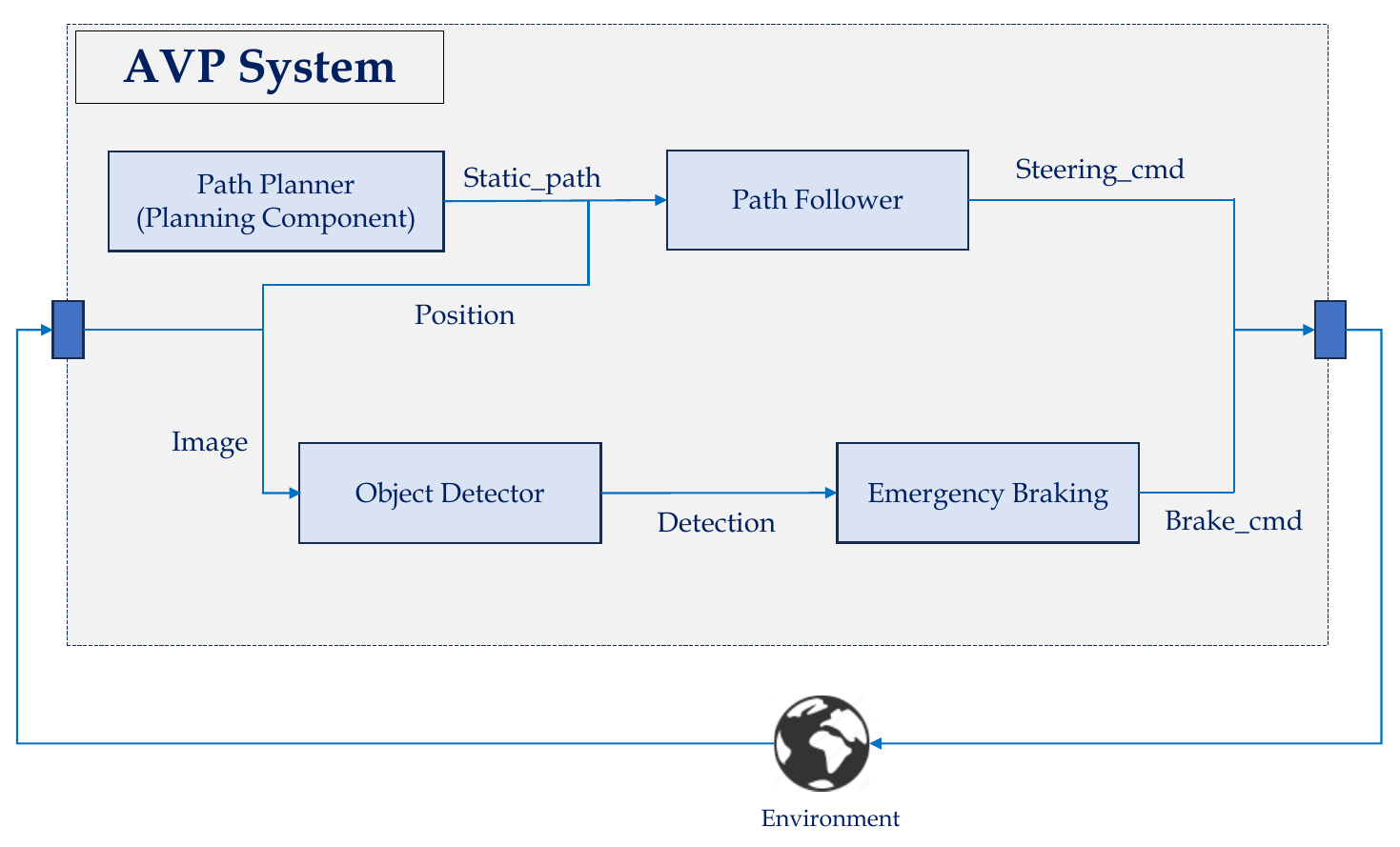}
    \caption{Overview of the architecture of the automated valet parking system.}
    \label{fig:avp-sut}
\end{wrapfigure}

The architecture of the system, which consists of several components, is shown in Fig.~\ref{fig:avp-sut}. We give a brief description of the components: The \textit{Planning Component} (PC) calculates a feasible and collision-free path for the automated vehicle given the locations of the drop-off zone and a designated parking spot. The \textit{Path Follower} (PF) controls at every time stamp whether the vehicle follows a pre-calculated static path. During this operation, the \textit{emergency brake} is triggered whenever there is an obstacle being recognized by the \textit{Object Detector} (OD) within a safety distance from the vehicle. The path follower and the emergency brake together form the \textit{Control Component} (CC) of the \AVP system.  To allow the save operation of \AVP system, the following main safety requirement should be fulfilled~\cite{esen2023simulationbased}: \textit{``\texttt{REQ}: The ego vehicle shall not collide with pedestrians, unless its velocity is zero."}. 


\textbf{Safety Assurance Challenges.}  However, maintaining an assurance case for \AVP system, which claims that \textit{\texttt{REQ}} is satisfied faces the following challenges:

(i) \textit{Dynamic Assurance}: The \AVP system contains a LEC, i.e. the OD component for the perception of other actors and objects. LECs are in general complex systems with large and multidimensional input spaces, whose correct behavior is difficult to be verified. In particular, it is not possible to know how these systems behave for any possible input when they are deployed on the street. Therefore, assuring the safety of a system whose decisions are based on LECs is complex. 
For that reason, a continuous engineering process is a common procedure in the automotive domain~\cite{bensalem2023continuous}, where the system gets updated even after deployment, when for instance more operational data is collected to retrain the LEC model of the OD to improve the models' performance. However, when systems components are updated, such as the OD, an updated assurance case has to be recreated. This recreation can be time and cost-intensive and requires a validation that all claims are considered for the assurance of \textit{\texttt{REQ}}. 

(ii) \textit{Tool Integration:}~Construction of an assurance case for \AVP involves the usage of different tools, encompassing formalization tools \cite{Bartocci23HyTem}, verification tools~\cite{infer,cppcheck, brian2023transformers}, or testing-related tools \cite{sorokin2023opensbt, Prescan}. In particular, the certification standard ISO 26262 in automotive, encourages the use of complementary tools when a single tool is not sufficient for a given test, analysis, or verification activity. The correct orchestration of these tools is manually possible but difficult due to the following reasons: 1) artifacts which that tools generate have to be manually maintained/tracked, 2) the exchange of data between tools is hard coded and have to be reimplemented from scratch, when tools are replaced by other tools.

(iii) \textit{Distribution}: ~Cyber-physical systems such as \AVP are in general based on different components that are developed by distinct organizational entities.  Maintaining an assurance case, requires the collection of evidence from different sources to construct an assurance case for the complete system. Also, employed tools can have different requirements on the infrastructure, they are deployed on which necessitates a distributed setup. For instance, testing tools require resource-intensive simulation environments, while verification tools may not.




Given theses challenges, an automated support for the continuous maintenance of the safety assurance throughout the lifecycle of \AVP is important to save operational costs and guarantee that all envisioned claims have been collected.

%% file: Sections/SX_Application.tex
\section{Assurance Case Development Using ETB}
\label{sec:application}

In this section, we illustrate the development of the assurance case for the \AVP system using the framework \ETB. \ETB is developed for the execution of distributed transactions and has been already applied to resolve the challenge of automating software certification workflows for the creation of assurance cases  \cite{cruanes2013tool,ruess2023evidential,shankar2022descert}. It enables an end-to-end, decentralized, and continuous safety assurance process where multiple entities are involved to establish safety claims supported by evidence. We outline how to use the framework \ETB to establish the assurance case for the \AVP system. Specifically,  we commence by introducing our assurance case pattern developed for the \AVP system. Then, we describe the formalization of the assurance case pattern in the language supported by \ETB, followed by the tool integration. Finally, we sketch the distributed creation and incremental maintenance of assurance cases.

\subsection{Assurance Pattern Creation}
\begin{figure*}[t]
\centering
\includegraphics[scale=0.42]{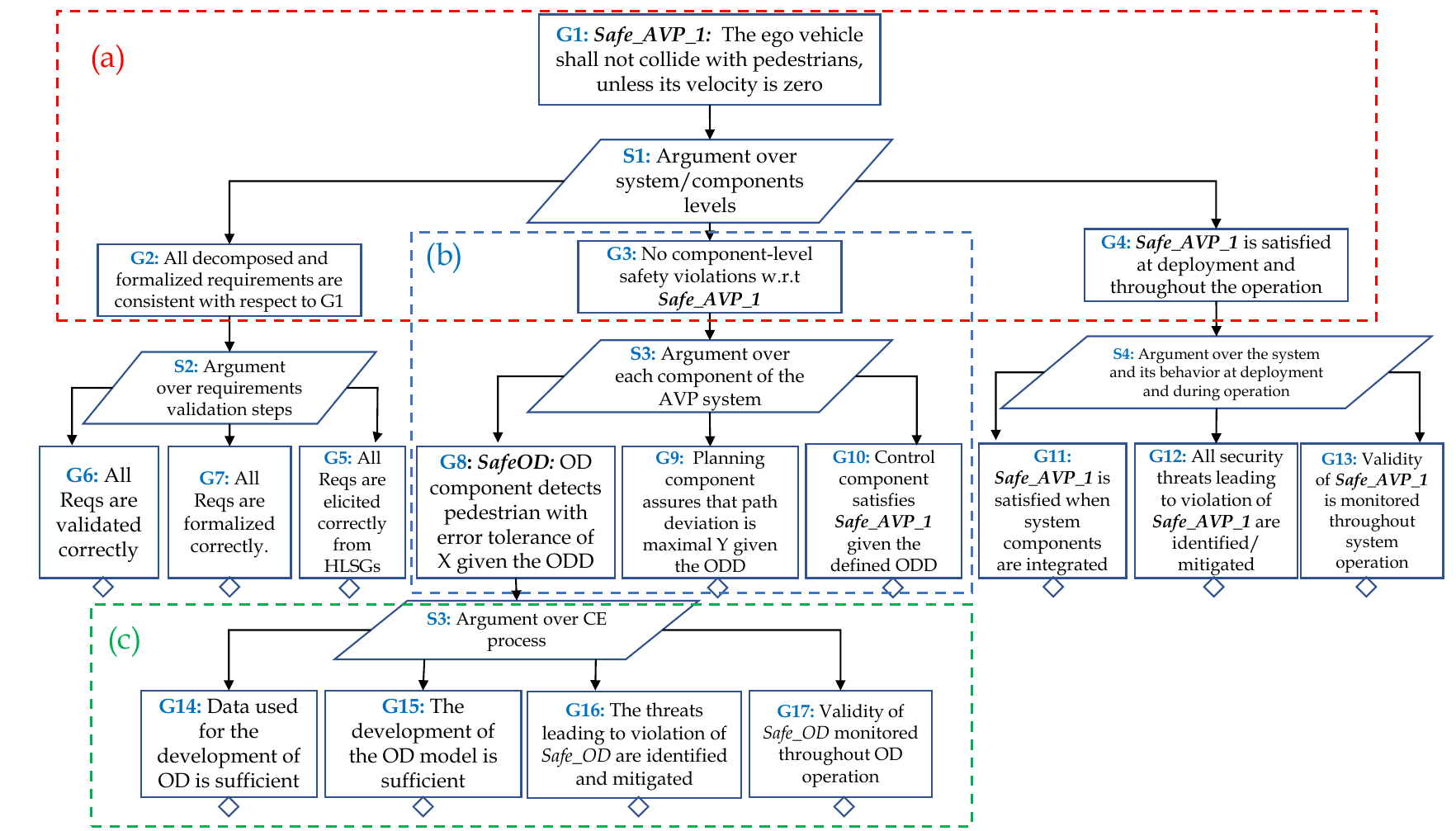}
    \caption{Top-level of GSN-based pattern for safety case construction for \AVP system.}
    \label{Fig:TopLevelGSN}
\end{figure*}

Let us consider a top-level assurance goal for the \AVP system related to the safety requirement \textit{\texttt{REQ}}. Although the primary goal here is not to develop a complete assurance case for the entire \AVP system, in Fig.\ref{Fig:TopLevelGSN}, we highlight three fragments of the argument pattern corresponding to various abstraction levels of the system with respect to the specified safety goal. 

The first argument pattern, which is shown in GSN-like \footnote{In this work, we have only considered the Goal and Strategy elements as well as the Supported-By relation of GSN in our assurance case fragments.} notation in Fig.~\ref{Fig:TopLevelGSN}(a), divides the top-level goal, which is labeled as $\goal_1$, into three sub-goals, $\goal_2$, $\goal_3$ and $\goal_4$, where $\goal_2$ targets the validation of the input requirements, and, $\goal_3$ and $\goal_4$ target the safety of the individual components and the overall system, respectively. The second argument pattern is dedicated to composing assurance arguments from each component of the \AVP system. 
As shown in Fig.~\ref{Fig:TopLevelGSN}(b), its three sub-goals, $\goal_8$, $\goal_9$ and $\goal_{10}$, target OD, PC, and CC of the \AVP system.  The third argument pattern, shown in Fig.~\ref{Fig:TopLevelGSN}(c), targets the OD component. Various recent approaches for the safety assurance of LECs propose safety assurance patterns over the LECs life-cycle \cite{hawkins2021guidance,schwalbe2020structuring,wozniak2020safety}. Building on such approaches, our argument pattern for the OD component targets stages such as data assurance, model assurance, verification assurance, and run-time assurance. As shown in Fig. \ref{Fig:TopLevelGSN}(c), its sub-goals, $\goal_{14}$, $\goal_{15}$, $\goal_{16}$ and $\goal_{17}$, target design-time (data, model and verification) and run-time assurance methods and evidence for the OD component. The three argument patterns illustrate how assurance patterns can be constructed for the \AVP system at different granularity levels by applying appropriate strategies. 

\subsection{Formalisation of Assurance Argument Patterns} 

As next, developed argument patterns should be specified as so called \textit{workflows} and formalized in Datalog, which is the scripting language supported by \ETB ~\cite{datalogCeri89}. For example, the top-level argument pattern (see Fig.~\ref{Fig:TopLevelGSN}) is formalized using the Datalog program that is given in Fig. \ref{Fig:DatalogAVP}. The head predicate is the top-level goal $\goal_1$, and its body contains each sub-goal of the argument pattern as a conjunct.

\begin{small}
\begin{figure*}[!h]
\centering
\lstset{style=prolog-style-small}
\begin{lstlisting}[label=datalog1]     
g1_safe_AVP(SUA, Reqs, ODD, Datasets, Specs,  RepSafOD, RepSafPC, 
            RepSafCC, ScenariosSBT, CrTests, RepCSM, RepME) :-
    subcomponents(SUA, [OD, PC, CC]),
    g2_reqs(Reqs, Specs),
    g3_safe_components([OD, PC, CC], Specs, ODD, Datasets, RepSafOD, RepSafPC, RepSafCC),
    g4_safe_system(SUA, Specs, ODD, ScenariosSBT, CrTests, RepCSM, RepME).
g2_reqs(Reqs, Specs) :-
    g6_req_validation(Reqs),
    g7_req_formalisation(Reqs, Specs).
g3_safe_components([OD, PC, CC], Specs, ODD, Datasets, RepSafOD, RepSafPC, RepSafCC) :-
    g8_safe_perception(OD, Specs, ODD, Datasets, RepSafOD),
    g9_safe_planning(PC, Specs, RepSafPC),
    g10_safe_steering(CC, Specs, RepSafCC).
g4_safe_system(SUA,Specs,ODD,ScenariosSBT,CrTests,RepCSM,RepME):-
    g11_scenario_based_testing(SUA,Specs,ODD,ScenariosSBT,CrTests),
    g12_cyber_security_monitoring(SUA, Specs, RepCSM),
    g13_monitoring_and_enforcement(SUA, Specs, RepME).     
\end{lstlisting}
\caption{Datalog formalization of the complete assurance pattern.}
\label{Fig:DatalogAVP}
\end{figure*}
\end{small}

Note, that the goal predicates in the Datalog rule contain all the variables the goal depends on and a descriptive identifier, and not just a goal index like the case for the assurance pattern. 
For instance, $\goal_1$ in the assurance pattern is formalized as the following Datalog predicate: 
\begin{center}

\lstinline+g1_safe_AVP(SUA, Reqs, ODD, Datasets, Specs,RepoSafOD, RepoSafPC,RepoSafCC, ScenariosSBT, CrTests, RepCSM, RepME)+\\
\end{center}

In this predicate, the string \lstinline+g1_safe_AVP+ is a descriptive post-fix of the goal name, and the parameters consist of the inputs such as \lstinline+SUA, Reqs, ODD, Datasets+ as well as resulting evidence artefacts like \lstinline+CrTests+, which is a set of considered to be critical test cases identified by the system-level testing method. We illustrate the description of the parameters used in the datalog formalization in table \ref{tab:decParDatalog}.

\begin{table}[!h]

\centering
\caption{Description of the parameters used in the Datalog formalization. \lstinline+I+ refers to input variables, \lstinline+O+ refers to output variables.}
\resizebox*{!}{1.5in}{
\begin{tabular}{|l|l|c||l|l|c|}
\hline
  \multicolumn{1}{|c|}{\textbf{Parameter}} &
  \multicolumn{1}{c|}{\textbf{Description}} &
  \multicolumn{1}{c||}{\textbf{I/O}} &
  \multicolumn{1}{c|}{\textbf{Parameter}} &
  \multicolumn{1}{c|}{\textbf{Description}} &
  \textbf{I/O} \\ \hline \hline
   \lstinline+SUA+ &
  System Under Assurance &
  \lstinline+I+ &
  \lstinline+RepSafPC+ &
  Report on safe PC &
  \lstinline+O+ \\ \hline
  \lstinline+Reqs+ &
  Requirements &
  \lstinline+I+ &
  \lstinline+RepSafCC+ &
  Report on Safe CC &
  \lstinline+O+ \\ \hline
  \lstinline+ODD+ &
  Operational Design Domain &
  \lstinline+I+ &
  \lstinline+ScenariosSBT+ &
  \begin{tabular}[c]{@{}l@{}}ScenariosSBT generated by system \\ based testing\end{tabular} &
  \lstinline+O+ \\ \hline
   \lstinline+Datasets+ &
  \begin{tabular}[c]{@{}l@{}}Datasets used for training, \\ testing, and validating OD\end{tabular} &
  \lstinline+I+ &
  \lstinline+CrTests+ &
  \begin{tabular}[c]{@{}l@{}}Critical test cases generated \\ by simulation-based testing\end{tabular} &
  \lstinline+O+ \\ \hline
   \lstinline+Specs+ &
  \begin{tabular}[c]{@{}l@{}}Formal specifications\\  of Reqs\end{tabular} &
  \lstinline+O+ &
  \lstinline+RepCSM+ &
  \begin{tabular}[c]{@{}l@{}}Report on cyber security \\ monitoring\end{tabular} &
  \lstinline+O+ \\ \hline
   \lstinline+RepSafOD+ &
  Report on Safe OD &
  \lstinline+O+ &
  \lstinline+RepME+ &
  \begin{tabular}[c]{@{}l@{}}Report on \AVP monitoring and \\ enforcement\end{tabular} &
  \lstinline+O+ \\ \hline
\end{tabular}
}

\label{tab:decParDatalog}
\end{table}

\subsection{Tool Integration} 
In this step, tools that provide evidence artefacts during the actual creation of assurance case are integrated into \ETB. 
Following the common practice in tool integration frameworks, 
\ETB provides a wrappers API that automatically generates wrapper templates that can be customized by end-users for each tool. 
%
%
As an illustration, the tool \texttt{OpenSBT}, designed for the search-based testing (SBT) of automated driving systems \cite{sorokin2023opensbt}, is offered as a service through the Datalog predicate \lstinline+g11_scenario_based_testing(SUA, Specs, ODD, ScenariosSBT, CrTests)+. In this predicate, \lstinline+ScenariosSBT+ and \lstinline+CrTests+ are artefacts generated by \texttt{OpenSBT} and represent respectively failing test inputs and corresponding simulation traces, which contain positions, the velocity of actors over time. 

In \ETB each tool can be evaluated under so-called one or more \textit{modes}. A mode specifies which variables serve as input or output of the corresponding tool. For example, the mode used for the predicate \lstinline+g11_scenario_based_testing(SUA, Specs, ODD, ScenariosSBT, CrTests)+ is \texttt{(+,+,+,-,-)}. That means that the first input argument, which is the \SUA, is of type string and holds the path to the system-under-assurance to be tested in the SBT tool, the second and third input arguments are files that capture the specifications and ODD constraints, while the two last arguments are files generated by \texttt{OpenSBT}. An example of a wrapper used to invoke \texttt{OpenSBT} is shown in Fig.~\ref{OpenSBTWRP}.   

\begin{figure}[!h]
\begin{lstlisting}[style=mystyle, language=java] 
public class OpenSBTWRP extends OpenSBTETBWRP {
  @Override
  public void run() {
    if (mode.equals("+++--")) {
      //1. Invoke OpenSBT  
      String suaPath = arg1;
      File specs = new File(arg2);
      File oddFile = new File(arg2);
      String odd = getOddFromFile(oddFile);
      String openSBT_CMD ="bash interface.sh "+ suaPath + " "+ specs + " "+ odd;
      String openSBT_Outputs = this.runCMD0(openSBT_CMD);
      //2. Evidence generation  
      String[] paths = openSBT_Outputs.split("\n");
      String criticalTcFilePath = this.workSpaceDirPath+"CriticalTC";
      this.createLocalFile(paths[0], criticalTcFilePath);
      this.arg3 = criticalTcFilePath;
      arg4 = new ArrayList<String>();  
      this.createLocalFiles(paths, arg4);
      //Add claim to claimDB
      this.addClaimPredicate();
    } 
    ...
}

\end{lstlisting}
\caption{Implementation of a wrapper for the integration of \texttt{OpenSBT} into \ETB.}
\label{OpenSBTWRP}
\end{figure}

\subsection{Distributed Assurance}



The usage of \ETB enables the creation of both assurance cases and evidence artefacts by executing the Datalog workflows in a distributed setting. Practically, a network of so called \ETB nodes can be defined where each node can create assurance cases or evidence artefacts depending on the type of the workflow it contains. In particular, each entity that wants to contribute to the assurance case - by providing a tool -  has to deploy an \ETB node. When a workflow is given to an \ETB node, it automatically identifies tools of other nodes which can contribute with their evidence to the overall assurance case.


As an example, consider the usage of a tool that helps to adapt the ODD at the runtime of the \AVP system. In particular, this tool derives at runtime of \AVP constraints from execution traces where the system behaves critically or not. Such tool can be for instance \texttt{HyTeM}~\cite{Bartocci23HyTem}. The derived constraints, i.e., the updated ODD, serve as an input for the system-level-testing tool \texttt{OpenSBT} (s. described in the previous section). While \texttt{OpenSBT}, is a testing tool that needs a comprehensive simulation environment with high resource usage, \texttt{HyTeM} requires considerably less resources and can be deployed in a different environment. The collection of distributed evidence provided by \texttt{HyTeM} and \texttt{OpenSBT} can be managed by~\ETB by the deployment of two \ETB nodes: where one \ETB node provides evidence from \texttt{OpenSBT}, while another \ETB node is deployed on a different platform.

\subsection{Automated Safety Case Creation}

\begin{figure*} [t]
\centering
\begin{subfigure}[t]{0.33\textwidth}        \centering
    \includegraphics[scale=0.27]{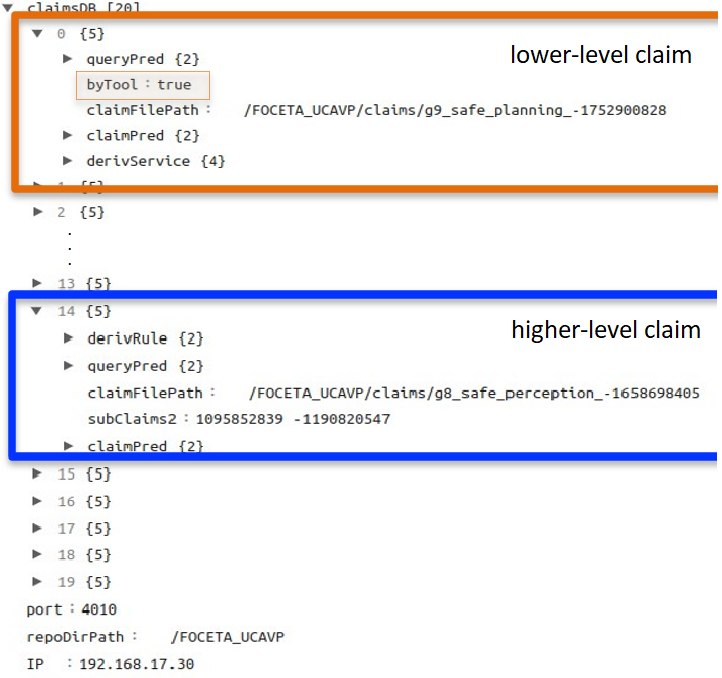}
    \caption{}
\end{subfigure}
\hfill
\begin{subfigure}[t]{0.6\textwidth}        \centering
    \includegraphics[scale=0.2]{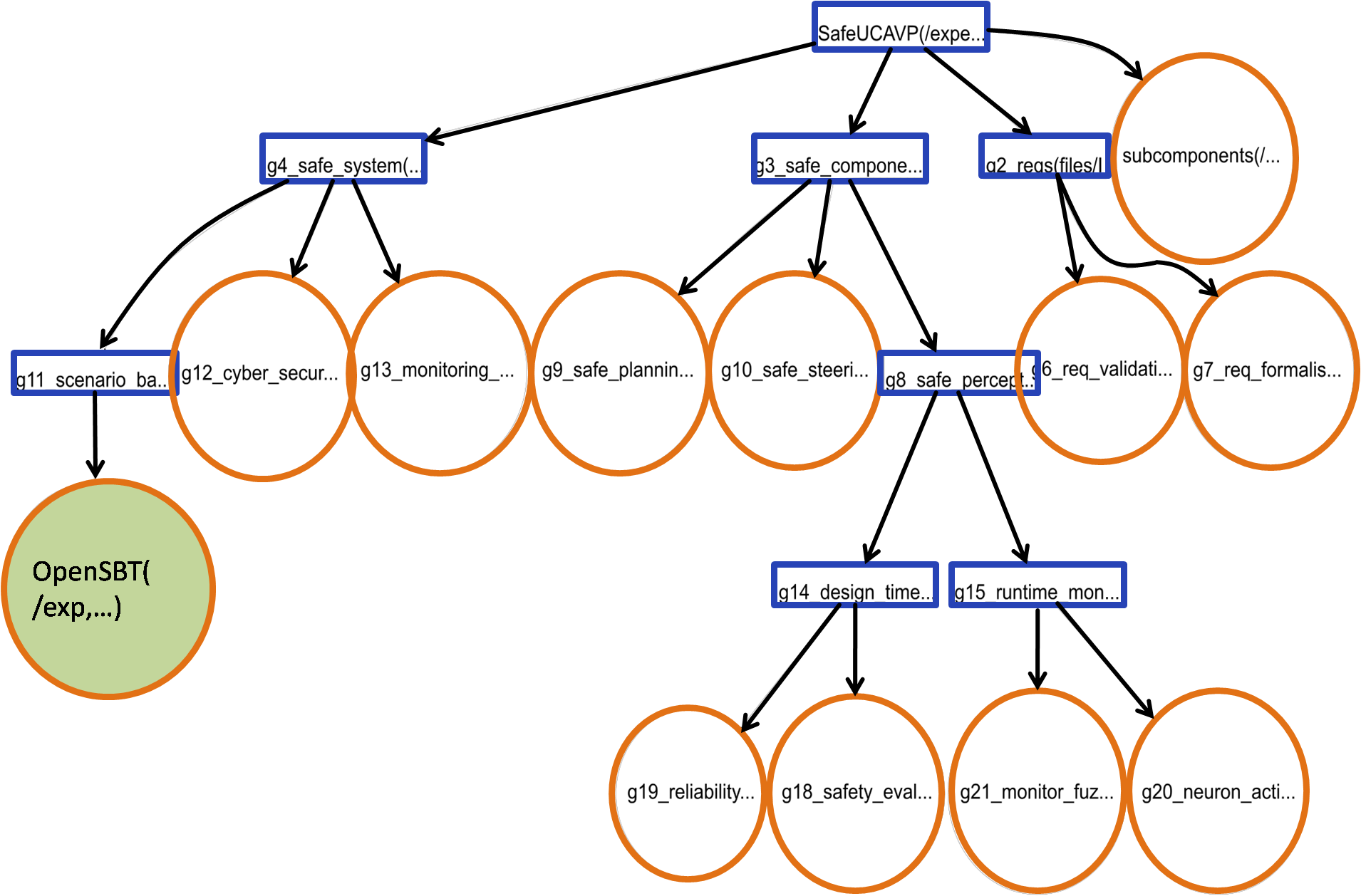}
       \caption{}
\end{subfigure}
\caption{a) shows an excerpt of the claim database of ETB in JSON format and b) shows an overview of corresponding claims in GSN-based format.}
\label{fig:claimDB}
\end{figure*}

\ETB provides a top-down left-to-right Datalog engine that automatically executes the assurance workflow specified in Datalog.
A datalog claim that can be instantiated, corresponds to an assurance claim with sufficient evidence. By the end of the execution of the assurance workflow, \ETB returns either the list of claims, i.e., a complete assurance case or a counter-example that points to the failed step (sub-goal or tool execution) of the assurance workflow. The established claims, sub-claims, and evidence, which compose the assurance case for the \AVP system, are given in Fig. \ref{fig:claimDB}. Fig. \ref{fig:claimDB} a) shows the claims in the database in \ETB and Fig.~\ref{fig:claimDB} (b) shows the corresponding GSN of the safety case. 


\begin{figure} [!h]
\hspace*{-1.2cm}
    \centering
    \includegraphics[scale=0.37]{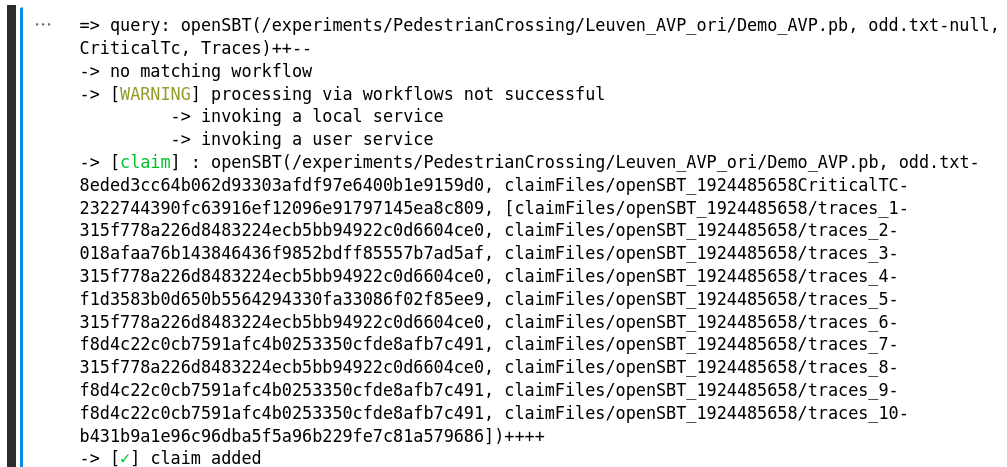}
    \caption{Console output of ETB after triggering OpenSBT. OpenSBT identifies critical test inputs with corresponding traces whose path is given on the right half of the Figure. These artefacts serve as evidence and are managed internally by ETB using a hash identifier as shown on the left part of the Figure.}
    \label{fig:OpenSBTEv}
\end{figure}

As visualized in Fig. \ref{fig:claimDB}(a), the created claims can be further categorized into two classes, namely \textit{high-level claims} and \textit{low-level claims}. While high-level claims correspond to the goals in the argument patterns that are supported by a set of sub-claims and a connecting argument, low-level claims are directly supported by evidence artifacts. For instance, the highlighted claim in green in Fig. \ref{fig:claimDB}(b) corresponding to the goal $\goal_{11}$, is supported by the evidence generated by \texttt{OpenSBT}, as depicted in Fig. \ref{fig:OpenSBTEv}. \ETB stores all these claims in a claims table and keeps track of all relevant and integrated tools and workflows w.r.t a given claim. 

\subsection{Assurance Case Maintenance}

One defining feature of the proposed process is continuous assurance case maintenance, where updates to the \SUA, its requirements, or verification plans are anticipated. In the event of such updates, outdated assurance artefacts including assurance cases and evidence artefacts are incrementally maintained, i.e., only a minimal set of maintenance actions are identified and executed.  The \ETB framework enables such assurance process by continuously executing relevant Datalog programs for outdated assurance cases and by continuously invoking verification tools for outdated assurance artefacts.

Let us consider the top-level claim \lstinline+g1_safeAVP+ created for the \AVP system. A common practice involves leveraging operational data to refine LECs, facilitating adjustments/improvements based on the evolving operational environment. Specifically, operational data plays a crucial role in refining the requirements and ODD \cite{tonk2021towards}. Yet, any updates to the requirements render the existing safety case obsolete, and the generation of new claims and evidence becomes imperative. \ETB applies its lightweight static dependency computation procedure to compute a sub-tree of the assurance case impacted by such update. The procedure is applied to the formalized assurance pattern that was executed during the establishment of the top-level claim.

\begin{figure} [t]
\hspace*{-1cm} 
    \centering
    \includegraphics[scale=0.40]{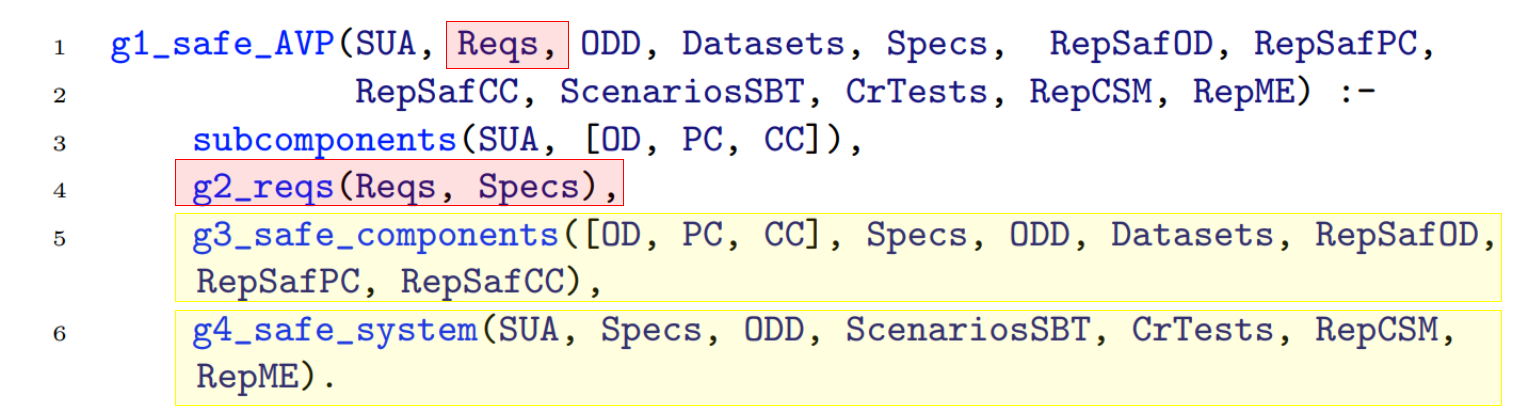}
    \caption{Example illustrating the incremental maintenance procedure after changing the
\lstinline+Reqs+ variable. Directly impacted goals and indirectly impacted goals are respectively marked in red and yellow.}
    \label{fig:SafetyCaseAVP}
\end{figure}

In particular, the maintenance procedure distinguishes between two types of goals: \textit{directly impacted goals} and \textit{indirectly impacted goals}~(s. Fig.~\ref{fig:SafetyCaseAVP}). Directly impacted goals, such as $\goal_1$, $\goal_2$, $\goal_6$ and $\goal_7$ are always re-run to re-establish their corresponding claims. However, indirectly impacted goals, which are $\goal_3$, $\goal_4$, $\goal_8$, $\goal_9$, and $\goal_{10}$, may not need re-running if the re-running of the directly impacted goals did not result in an update to their assurance artefact.

%% file: Sections/SX_Lessons.tex
\section{Discussion}
\label{sec:discussion}

In this section, we discuss on the experience of using the Evidential Tool Bus for assurance case maintenance and the limitations of \ETB, when generating assurance cases for automated driving systems.
The case study reveals that the \ETB framework provides computer-aided support for developing assurance cases, where it enables the automated execution of assurance workflows, fostering scalability in assurance evidence generation and maintenance. Additionally, \ETB facilitates the distributed execution of assurance workflows, along with the orchestration and integration of various tools utilized in evidence generation. 

However, we note that the specification, and validation of these Datalog-based workflows have to be done manually by the user. 
From our experience, the concrete manual specification of assurance workflows is manageable for simple workflows, as Datalog is a declarative specification language with a simple syntax. However, a comprehensive assurance case may involve a significant number of nodes to cover further aspects of the system's functionality such as fault tolerance, and to provide a sufficient analysis. Further, in our example, we did not prove the validity of single claims, such as, e.g., how confident the system-level testing tool \texttt{OpenSBT} is that a test case is critical or that no further failing tests exist. 
However, confidence scores can be similarly incorporated into the workflow specification and modeled as an argument in a Datalog predicate of the corresponding tool.



%% file: Sections/S2_RelatedWork.tex
\section{Related Work}
\label{sec:related}

This section provides a brief overview on research works on assurance cases and dynamic assurance case maintenance~\cite{wargContinous19}.
Recently, there has been work on assurance case pattern selection \cite{hawkins2021guidance,schwalbe2020structuring}, and assurance case pattern instantiation~\cite{kaur2020assurance,wozniak2020safety,RamakrishnaAutoGSN20,denney2018tool}.  We focus on the latter one, as this is the main challenge considered in our paper. 

Hawkins et al.~\cite{hawkins2021guidance} have presented a methodology for the instantiation of assurance cases for autonomous systems containing ML components. The methodology comprises a set of safety case patterns and a process for (1) systematically integrating safety assurance into the ML component life cycle, as well as (2) generating the evidence base for explicitly justifying the acceptable safety of integrated ML components. While AMLAS covers several stages of the ML life-cycle, computer-aided support to automate the assurance case construction as well as its maintenance in case of requirements or system changes, as supported by \ETB, is not discussed. In contrast, a recent work~\cite{shankar2022descert} proposed to ``automate" the evaluation of software assurance evidence to enable certifiers to determine rapidly that system risks are acceptable. They adopted a tool-based approach to the construction of software artifacts that are supported by rigorous evidence. This concept is very important, especially for certification, where it is desirable that arguments representing safety assurance can be re-playable.

Ramakrishna et al.~\cite{RamakrishnaAutoGSN20} developed the tool ACG for the automated assurance case generation, given a manually curated evidence store. The evidence store is populated with evidence artifacts, which are automatically generated from the system model architecture. While ACG automatizes the safety case instantiation, it lacks a mechanism to support dynamic safety assurance as enabled by \ETB. The toolset AdvoCATE~\cite{denney2018tool} supports the development of assurance cases and has been applied to a use case for swift unmanned aircraft system but also here no assurance case maintenance is possible.









%% file: Sections/S6_Results_Conclusion.tex
\section{Conclusion}
\label{Sec:Conclusion}

In this paper, we presented the utilization of the \ETB framework as computer-aided support for the development of an assurance case for an automated driving system. We illustrated the formalization of the assurance case as a Datalog program and have shown the integration of a testing tool to provide evidence for the argumentation of the safety of the system at the system level. In addition, we outlined the distributed execution of tools integrated with \ETB to cope with different levels of interoperability of tools and heterogeneity of providers, as well as described how the incremental assurance is supported by \ETB.

In our future work, we plan to extend our study incorporating all tools required to instantiate a complete assurance case for the \AVP system. Further, we are working on an approach that enables to generate ETB workflows, i.e., Datalog specifications, from assurance case patterns represented as GSN to facilitate the application of ETB for the assurance case creation. Furthermore, we plan to extend \ETB to support a comprehensive confidence argumentation \cite{hawkins2011new}. 


